\documentclass{aastex}
\usepackage{spr-astr-addons}
\usepackage{url}
\usepackage{adjustbox}
\usepackage{array}
\usepackage{verbatim}
\usepackage{stmaryrd}
\usepackage{epsfig}
\usepackage{epstopdf}
\usepackage{multirow}
\usepackage{booktabs}
\usepackage{rotating}
\usepackage{etoolbox}

\usepackage{boondox-cal}

\newcolumntype{L}[1]{>{\raggedright\let\newline\\\arraybackslash\hspace{0pt}}m{#1}}
\newcolumntype{C}[1]{>{\centering\let\newline\\\arraybackslash\hspace{0pt}}m{#1}}
\newcolumntype{R}[1]{>{\raggedleft\let\newline\\\arraybackslash\hspace{0pt}}m{#1}}

\RequirePackage{color}

\begin{document}

\title{Lepton capture rates due to isotopes of vanadium in astrophysical environment}
\shorttitle{Lepton capture rates of vanadium isotopes}
\shortauthors{Author et al.}

\author{Ramoona Shehzadi\altaffilmark{1}} \and \author{Jameel-Un Nabi\altaffilmark{2}} \and
\author{Fakeha Farooq\altaffilmark{1}}


\altaffiltext{1}{Department of Physics, University of the Punjab,
Lahore, Pakistan.} \altaffiltext{2}{Faculty of Engineering Sciences
(FES),\\GIK Institute of Engineering Sciences and Technology, Topi
23640, Khyber Pakhtunkhwa, Pakistan.} \altaffiltext{1}{Corresponding
author email : ramoona.physics@pu.edu.pk}

\begin{abstract}
Lepton (electron and positron) capture rates on iron-regime nuclei are an
essential element for modeling the late stages of progression of massive
stars that become core collapse and thermonuclear supernova. As per previous
simulation studies, lepton capture (LC) rates on isotopes of vanadium are
believed to have a substantial effect in regulating the Y$_{e}$ (lepton to
baryon fraction) during the final evolutionary phases. The present work involves
the calculation of lepton capture rates for 22 isotopes of vanadium by making use of
the proton-neutron (pn-) quasiparticle random phase approximation (QRPA)
model. The covered mass range is from A = 43 to 64. The LC rates have been computed
over stellar densities ranging from $10^{1}$ to $10^{11}$ (g/cm$^{3}$) and
for the temperature range $10^{7} - 3 \times 10^{10}$ (K). A comparison of
our calculated LC rates to the rates computed using other models (IPM and
LSSM) has also been presented. As compared to the rates calculated by other
models, pn-QRPA rates at high temperature ($3 \times 10^{10}$\;K) are
larger by 1-2 orders of magnitude.
\end{abstract}


\section{Introduction}
\label{sec:intro} Massive stars make headway towards core-collapse
(Type II) supernova by passing through hydrostatic and explosive
nuclear burning phases~\citep{Burbidge57, Bethe90, Rauscher02,
Busso99}. In the hydrostatic phase, successive burning of each layer
of the star's core contracts the core. Consequently, temperature,
density and mass of the core increases. At enough high densities,
under the Chandrasekhar mass limit ($\sim 1.5$ M$_\odot$), electron
degeneracy pressure resists further shrinking of the core and
battles against its collapse. Beyond the Chandrasekhar limit,
weak-decay processes, namely; $\beta$-decays and lepton captures
significantly affect the star's life cycle. Especially the lepton captures (LC), including electron captures
(EC) and positron captures (PC), have considerable role in the
pre-supernova phases of stellar development~\citep{Bethe79, Nabi07}.

During the late evolution stages, star's atmospheric temperature
approaches around $\sim 10^{9}\;$K, which leads to a considerable rise
in the Fermi energy, $E_{f}$, of the degenerate gas. This favors the capture of electrons
to heavy nuclei and free protons. The capturing processes
lower the electron density within the core and become a cause to
reduce Y$_{e}$ (lepton to baryon ratio) and produce neutrinos. For
matter densities up to $\sim 10^{11}$\;g/cm$^{3}$, these neutrinos
escape the star taking away energy and result in reduction of the
entropy of the core~\citep{Heger01}. Thus, the core
entropy, Y$_{e}$, and Chandrasekhar mass (proportional to Y$_{e}$)
are controlled by the LC processes~\citep{Lang03}. These factors
mainly decide the dynamical behavior of the collapsing of the
core. During the final evolutionary pre-supernova
stage, the core is thrust towards collapse. LC rates thus play a
decisive part in the modeling of Type II supernova as well as in the
dynamics of bounce shock~\citep{Hix03}. Primarily, EC rates are
requisite for the simulation of dynamical evolution of
supernovae. In addition, PC rates are
of fundamental significance in the stellar environment~\citep{Suz16, Nabi07}.
PC weak rates are appreciable at low density (in the range of
$10-10^{6}$\;g/cm$^{3}$) and high temperature regions. On contrary,
for high density (especially at relatively low temperature) EC wins
over PC~\citep{Nabi17}.

LC weak processes are mainly ruled by the Gamow-Teller (GT)
transition strengths in the late stages of star's evolution. GT
properties of Fe-peak nuclei are therefore of primary importance in
supernovae~\citep{Fuller}. An accurate determination of the weak
rates needs detailed knowledge of GT transition strengths in EC
(GT$_{+}$) and PC (GT$_{-}$) directions. The work of Fuller et al.,
(FFN)~\citep{Fuller},  which was based on the independent particle
model (IPM), reveals the importance of capture processes to the GT
resonance and they calculated capture rates for a wide domain of
temperature $10^{7}-10^{11}$ (K) and density $10^{1}-10^{11}$
(g/cm$^{3}$) for 226 nuclei with masses in the range 21 to 60. For
nuclei, with mass numbers greater than 60, Aufderheide et
al.,~\citep{Aufder96} updated the work of FFN with the consideration
of GT quenching. In a study, El-Kateb et al.,~\citep{Kateb94}
described the inaccurate placement of GT centroid in the
computations done by FFN~\citep{Fuller} and Aufderheide et
al.,~\citep{Aufder96}. Since, in the massive stars, with intensely
hot and dense environment, transitions from the excited states are
also notable besides the transitions from the ground states.
Therefore, an accurate estimation of the capture rates demands a
microscopic calculation of the ground, as well as the excited state
GT$_{\pm}$ strength functions. Two microscopic theories were
developed, the proton-neutron quasi-particle random phase
approximation (pn-QRPA) theory~\citep{Nabi99} and the large-scale
shell model (LSSM)~\citep{Lang00}. These theories are considered to
be most effective for an accurate estimation of stellar week rates.
The pn-QRPA model uses a large model space of 7$\hbar\omega$ for the
calculations. Moreover, unlike the LSSM, this model is not based on
Brink's hypothesis~\citep{BAH}, and calculates the GT strength
distributions by considering the contributions of all the excited
states of parent nuclei separately. These features enhance the
reliability of the pn-QRPA theory for computing the stellar week
rates. The reliability and validity of the pn-QRPA model could be
evaluated from Ref.~\citep{Nabi04}.

The pn-QRPA theory was employed for the first time by Nabi and
Klapdor-Kleingrothaus~\citep{Nabi99a, Nabi04} for
the computation of the weak decay rates of 709 nuclei (mass number
ranging 18-100) for a wide range of stellar temperatures and densities. Later, Nabi and collaborators, further improved the
algorithm by refining the model parameters and incorporating the
up-to-date experimental data for the calculation of the rates e.g.,
in~\citep{Nabi05, Nabi07, Nabi08}.

In the pre-supernovae development of high-mass stars, weak decay properties of several isotopes
of vanadium play an important part~\citep{Heger01, Aufder94}. Simulation studies of Aufderheide and collaborators show that $\beta$ decay of $^{50,52-57}$V and electron captures on
$^{50-55}$V are of significant importance as the product of their weak rates
and abundances can affect time rate of change of Y$_{e}$ substantially. Most
of the previous studies have focused mainly on the calculations of
weak decay properties of stable isotopes ($^{50}$V and $^{51}$V).
For example, in Ref.~\citep{Nabi07}, the authors discussed the
impact of GT strength distributions on EC rates on $^{50}$V.
Later,~\citep{Muneeb13} calculated the EC rates on $^{51}$V by
applying the pn-QRPA theory. The work of~\citep{Cole12},
~\citep{Sarrig13} and~\citep{Sarrig16} also highlighted the
importance of $^{50}$V and $^{51}$V in astrophysical environment. As future reserach may reveal further
astrophysically important vanadium isotopes, it motivated us to calculate the weak decay rates
for isotopes of vanadium other than $^{50}$V and $^{51}$V. Recently,
in a study~\citep{shehzadi20}, we have reported a detailed study
of the GT strength distributions, gamma heating rates
($\lambda_{\gamma}$)and neutrino cooling rates ($\lambda^{\nu}$) for
a chain of 22 isotopes of vanadium ($^{43-64}V$) by making use of the
deformed pn-QRPA theory. The terrestrial $\beta$-decay half-lives of
these isotopes were also estimated and were found to be in good
accordance with the experimental ones taken from~\citep{Aud17}. The pn-QRPA
deduced B(GT) data was also compared with the ones from other models
calculations, as well as the experimental data wherever available. A
comparison of our computed cooling rates with the ones calculated
using the IPM model and LSSM was also presented.

In the current work, we will focus on the computation of the lepton
capture weak rates of vanadium isotopes using our recently published
B(GT) data. The estimated LC rates are also compared with the previously
reported rates of IPM and LSSM. The adopted formalism based on
deformed pn-QRPA theory is briefly described in the next section.
Section~\ref{sec:results} presents the results and discussion on the
calculated LC rates of vanadium isotopes.
Section~\ref{sec:conclusions} covers the conclusions made on the
present work.

\section{Formalism}
\label{sec:formalism}

The hamiltonian implemented for the deformed pn-QRPA theory has the
form
\begin{equation}
\mathsf{H^{QRPA}} = h^{sp} + \nu^{pair} + \nu_{GT}^{pp} + \nu_{GT}^{ph} .
\end{equation}
here $h^{sp}$ stands for the single-particle ($sp$) hamiltonian,
whose energies and state vectors were estimated by using the
Nilsson model~\citep{Nil55}. $\nu^{pair}$ is the pairing
interaction, handled within the BCS approximation, and
$\nu_{GT}^{pp}$ is the particle-particle ($ph$) GT force and $\nu_{GT}^{ph}$ is
the particle-hole ($ph$) GT force. These forces
incorporate the residual interactions for proton-neutron pairs. The
force constants for $pp$ and $ph$ interactions were specified by
$\kappa$ and $\chi$, respectively. These constants are parameterized
as in~\citep{Hir93}
\begin{equation*}
\chi = 23/A \;(MeV);~~~~ \kappa = A^{-2/3} \;(MeV).
\end{equation*}
The same parametrization was used in our recent calculations of
B(GT) strength distribution functions and the energy rates of vanadium
isotopes~\citep{shehzadi20}. Among other required parameters of the
model are; the Nilsson potential (NP) parameters, which are adopted
from Ref.~\citep{Nil55} and the Nilsson oscillator constant was
taken as
\begin{equation}
\Omega = 41/A^{1/3}\;(MeV).
\end{equation}
The pairing gaps were calculated using the following relation,
\begin{equation}
\Delta _{n} = \Delta _{p} = 12/\sqrt{A}\;(MeV).
\end{equation}
The nuclear quadrupole deformation $\beta_{2}$, given by
\begin{equation}
\beta_{2} = \frac{125(Q)}{1.44(A^{2/3})(Z)} \label{Eq:deform}~,
\end{equation}
$Q$ is the electric quadruple moment whose values were adopted from
~\citep{Mol95} and A and Z are atomic mass and atomic number,
respectively. The Q-values as calculated in Ref.~\citep{Aud17} were taken.

The basic approach for the calculations of weak decay rates is based on
the formalism adopted in the prior work of FNN~\citep{Fuller}.
However, in our calculations a microscopic way is adopted to compute
the GT strength for capture rates from all parent excited levels.
The rates of the lepton capture, from parent $n^{th}$ state to the daughter
$m^{th}$ state is given as
\begin{eqnarray}
\lambda ^{LC} _{nm} &=& \left(\frac{\ln 2}{D} \right)
[f_{nm} (T, E_{f}, \rho)][B(F)_{nm} \nonumber \\
&+&(g_{A}/ g_{V})^{2} B(GT)_{nm}];~~~~ L\equiv E,P,\label{Eq:LC}
\end{eqnarray}
In Eq.~\ref{Eq:LC}, B(F)$_{nm}$ and B(GT)$_{nm}$ are known as
reduced Fermi and GT transition probabilities, respectively and were
determined as in~\citep{shehzadi20}. For present calculations, D =
6143~\citep{Har09} and the ratio $g_{A}/g_{V}$ is
-1.2694~\citep{Nak10}. The integral taken over total energy,
$f_{nm}$, is the phase space integral (natural units:
$c = \hbar = m_{e} = 1$ were used in its calculations). For lepton capture rates it is given as
follows (with upper sign for EC and lower for PC)
\begin{equation}
f_{nm} = \int _{\mathcal{w}_{l} }^{\infty }\mathcal{w}
(\mathcal{w}^{2} -1)^{1/2}(\mathcal{w}_{m} +\mathcal{w})^{2} F(\pm
Z, \mathcal{w}) G_{\mp }d\mathcal{w}. \label{phcapture}
\end{equation}
In this equation, $\mathcal{w}$ is total energy (sum of
kinetic and rest mass energies) of electron or positron, $\mathcal{w}_{m}$
is the total energy of $\beta$-decay and
$\mathcal{w}_{l}$ is the total threshold energy for the lepton capture.
The G$_{\mp}$ are the Fermi-Dirac lepton distribution function.
The Fermi functions  $F\left(\pm Z,\mathcal{w}\right)$ are
calculated by adopting the method described in~\citep{Gove71}. The total
lepton capture rates are given by;
\begin{equation}
\lambda^{LC} = \sum _{nm}P_{n} \lambda _{nm}^{LC};~~~~ L\equiv E,P,
\label{rate}
\end{equation}
Here $P_{n}$ is the occupation probability of parent excited levels,
which obeys the normal Boltzmann distribution. The summation in
above equation was taken over set of all parent and daughter nuclei
states until the calculated rates were converged.

\section{Results and discussions}
\label{sec:results}

In this analysis, we have calculated lepton capture rates for twenty two
isotopes of vanadium using deformed pn-QRPA model. The mass of nuclides
ranges from A = 43 to A = 64. The list includes the unstable isotopes
including the neutron rich isotopes ($^{43-49}$V and $^{52-64}$V), as well as the
stable ones ($^{50}$V, $^{51}$V). The calculations have been done for the
stellar temperatures ($10^{7} - 3 \times 10^{10}$)\;K and density in the
range ($10-10^{11}$)\;g/cm$^{3}$. Our calculated rates have also been
compared with the previous calculations done using LSSM and IPM. The
calculated lepton capture rates due to $^{43-64}$V isotopes are given in
Tables~\ref{V43-46}-\ref{V62-64}. For space limitations, we have presented
the rates only at some selected values of density and temperature. The
$1^{st}$ column of the tables depicts the values of $\log\rho \text{Y}_{e}$ having
units of g/cm$^{3}$, where $\rho$ is the baryon density. In the
second column, stellar temperature (T$_{9}$) are stated in units of $10^{9}$\;K.
The remaining columns give the pn-QRPA estimated lepton capture
rates for our list of vanadium isotopes in units of s$^{-1}$.
$\lambda^{\text{EC}}$ ($\lambda^{\text{PC}}$) denotes the weak decay rates because of electron capture (positron capture). As can be observed from
Tables~\ref{V43-46}-\ref{V62-64}, the values of EC rates rise as the stellar
temperature and density increase. It happens as the density of the stellar core
increases, the Fermi energy, $E_f$ of the electrons rise which cause an
enhancement in electron capture rates. Additionally, an increase in the
stellar temperature results in an enhanced occupation probability of excited
states of parent nuclei which result in an effective contribution to the
sum total of weak decay rates. From tables, it can also be observed that, with the increasing temperature, there is a reduction in the rate
of change of EC rates in the high density region
($\log\rho \text{Y}_{e}$ = 11). Tables show that, as the stellar temperature
rises, the positron capture rates increase. However, in contrast to
EC rates, with increasing core density, there is reduction in
PC rates. As the core density decreases or the temperature rises (the
degeneracy parameter for positrons at this stage becomes negative),
high-energy positrons are generated resulting in an increase of positron capture rates. It
can also be noted from Tables~\ref{V43-46} and~\ref{V47-50}, that for
$^{43-50}$V isotopes, PC rates are smaller as compared to the corresponding
EC rates by quite a few orders of magnitude. For isotopes having $A \ge 51$, at
low and medium densities $(\log \rho$Y$_{e}$ = 3 and 7), for temperatures
(T$_{9} \le 10$), PC rates become bigger than the corresponding EC rates. For
higher stellar temperature (T$_{9}$ = 30), the electron capture rates for
$^{51-57}$V isotopes become comparable to the PC rates.
However, for $A > 57$, as we go to more neutron abundant isotopes, the PC rates
are still larger in comparison to the corresponding EC rates by 1-2 orders of magnitude.
For high density domain, the EC rates prevail the PC rates and hence the PC
rates can be ignored (specially at low temperatures; T$_{9} < 30$). The rates at fine temperature-density grid are available and may be requested from the corresponding author.

\begin{figure}
\begin{center}
\includegraphics[width=0.8\textwidth]{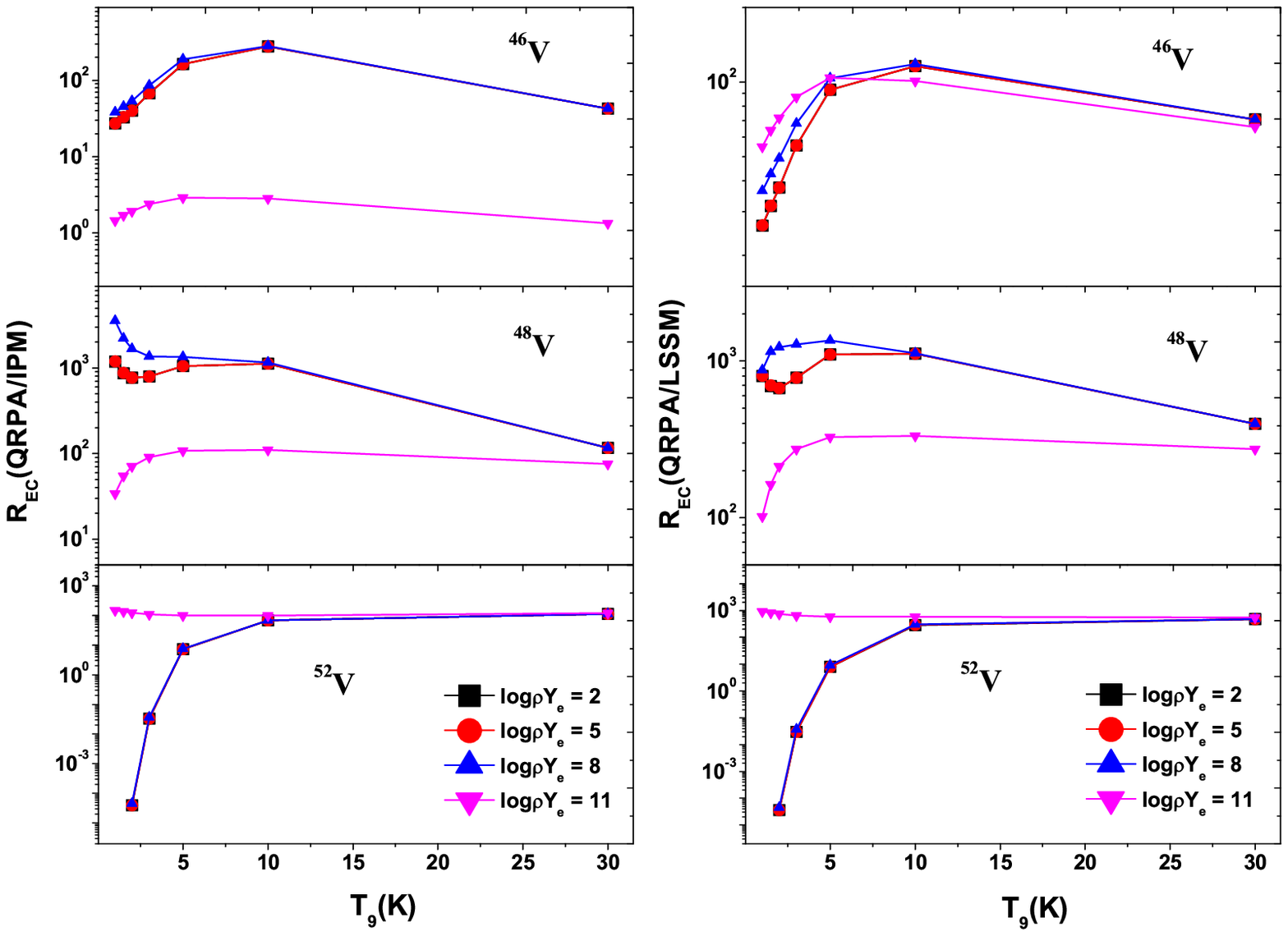}
\includegraphics[width=0.8\textwidth]{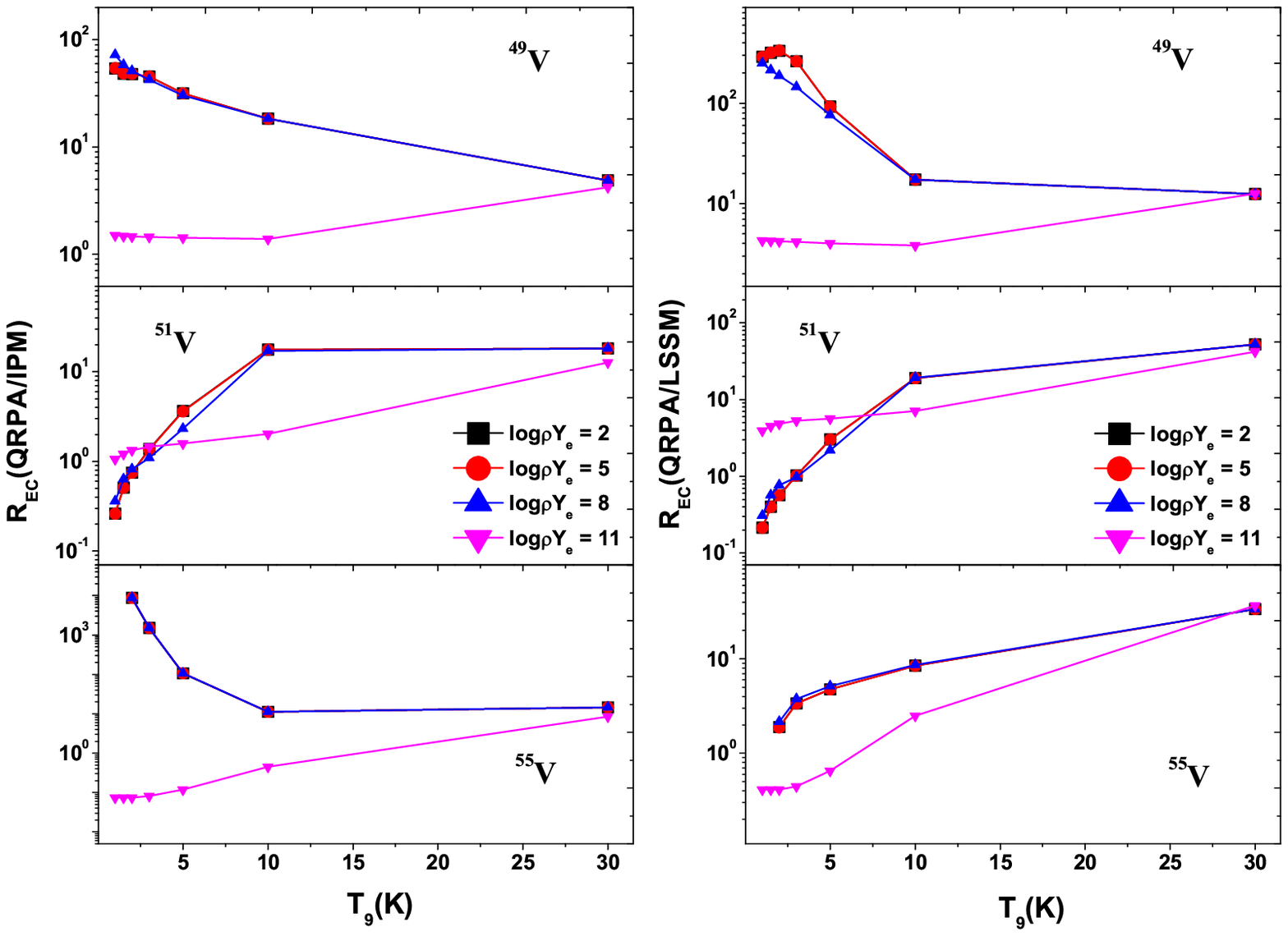}
\vspace{-0.1cm}\caption{The comparison of pn-QRPA calculated
electron capture rates due to some of even-A (odd-A) vanadium
isotopes at top (bottom) with the previous calculations performed by
IPM (on left) and LSSM (on right) at different selected densities as
a function of stellar temperature. $\log\rho$Y$_{e}$ gives the
$\log$ to base 10 of stellar density in units of g$\;$cm$^{-3}$.}
\label{EC_Comp}
\end{center}
\end{figure}

\begin{figure}
\begin{center}
\includegraphics[width=0.8\textwidth]{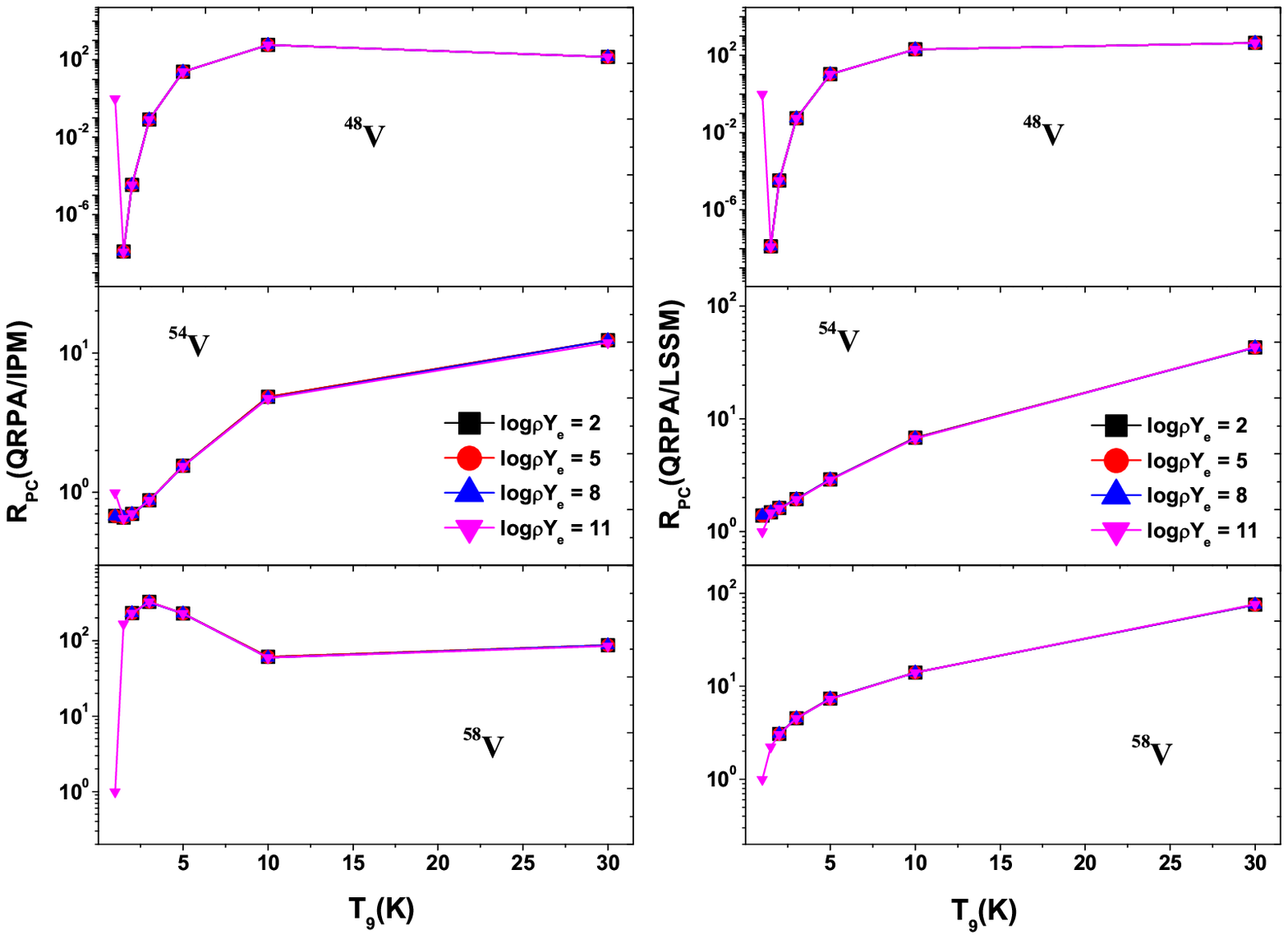}
\includegraphics[width=0.8\textwidth]{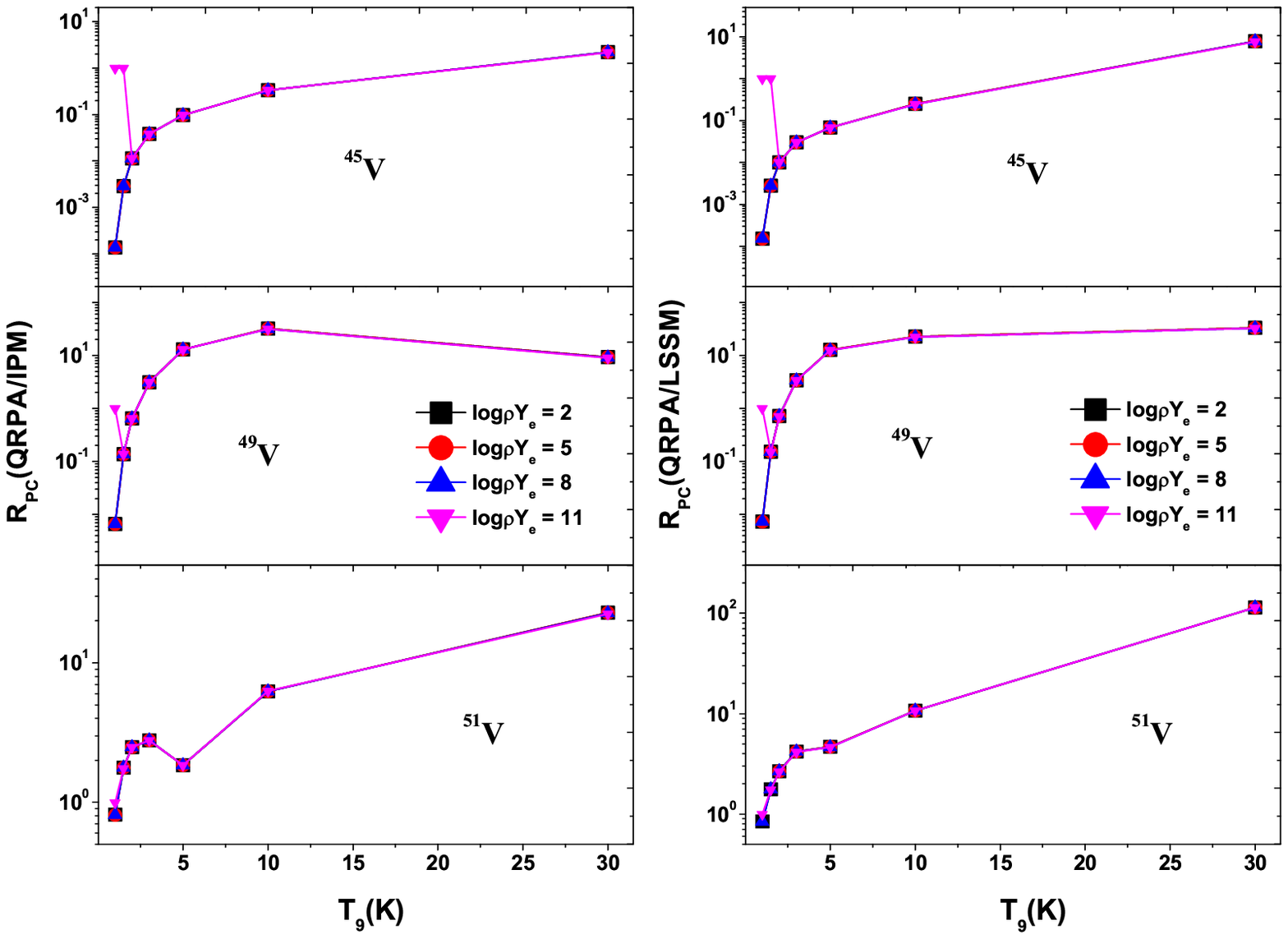}
\vspace{-0.1cm}\caption{The comparison of pn-QRPA calculated
positron capture rates due to some of even-A (odd-A) vanadium
isotopes at top (bottom) with the previous calculations performed by
IPM (on left) and LSSM (on right) at different selected densities as
a function of stellar temperature. $\log\rho$Y$_{e}$ gives the
$\log$ to base 10 of stellar density in units of g$\;$cm$^{-3}$.}
\label{PC_Comp}
\end{center}
\end{figure}

Next, we present the comparison of our estimated LC rates to
those of calculated by LSSM and IPM. We have calculated ratios of our
calculated lepton rates to the corresponding rates computed by LSSM and IPM,
which have been presented in the form of graphs (see Figures~\ref{EC_Comp}
and~\ref{PC_Comp}). The ratios (along y-axis) have been plotted against
stellar temperature at four different values of density ($\log \rho$Y$_{e}$ =
2, 5, 8, 11). The left panel of the graphs shows the ratios of our
calculated EC/PC rates to the corresponding IPM rates and in the right panel
the comparison with respect to LSSM rates has been presented. We had observed
that the comparison graphs of several isotopes showed a similar trend and
hence we have presented the comparison results only for a few selected cases due
to the space consideration. Figure~\ref{EC_Comp} shows the comparison of EC rates
for three selected even-A (odd-A) isotopes, $^{46, 48, 52}$V ($^{49, 51,
55}$V) at top (bottom). In case of $^{45-49,53,56}$V, QRPA-calculated rates
are in general greater than the corresponding IPM and LSSM rates at all values of
temperature and density. For example, in case of $^{46,49}$V ($^{48}$V), our
computed EC rates are bigger than the rates calculated by IPM and LSSM by 1-2
(1-3) orders of magnitude (see Figure~\ref{EC_Comp}). This difference in the
EC rates estimated using pn-QRPA theory and the other models could be
attributed to their different way of computation of GT transition strength
contribution from the excited levels. Brink's hypothesis was applied in the
calculations of IPM and LSSM for the estimation of excited states GT
transition strengths. This hypothesis is based on the assumption that the
contribution to GT strengths from the excited levels is comparable to that from
the ground state, with the exception that the earlier ones are shifted by the
excitation energy of the states. In contrast to this approach, pn-QRPA model
adopted a microscopic way of computation of GT transition
strength contribution from all the parent's exited levels separately. The summation was
then applied over all sates of daughter and parent nuclei to evaluate the
total capture rates.

In case of $^{50-52}$V, at lower temperatures and densities, LSSM and IPM EC
rates are in general bigger than the corresponding pn-QRPA rates. However, at
higher temperatures ($T_{9} > 3$) and density, our rates surpass the rates
calculated using LSSM and IPM. One of such instance from odd-A nuclei
($^{51}$V), and one from even-A nuclei ($^{52}$V), is shown in
Figure~\ref{EC_Comp}. For $^{51}$V, pn-QRPA-computed EC rates are bigger as compared to
the IPM (LSSM) rates by up to a factor of 18 (52) at higher temperatures. In
case of $^{52}$V, at higher temperature and density our estimated rates surpass the IPM
and LSSM rates by up to $~\sim$2 orders of magnitude. In the last row of
Figure~\ref{EC_Comp} the rates of $^{55}$V are shown. For this isotope,
except at high density and low temperatures, our rates are bigger. In
addition to the earlier stated different way of calculation of excited level
GT strength distribution, some other reasons could contribute to the
differences in our and other model's calculated rates. For instance, the
calculations of IPM used wrongly placed GT centroid, didn't consider
quenching and also suffered from the approximations applied in the
calculations of nuclear matrix elements. In case of calculations done by
LSSM, some convergence issues were observed as stated in
Ref.~\citep{Pruet03}. pn-QRPA theory did not face any such problems and
calculates rates in a microscopic manner.

Finally, we move to the comparison of PC rates calculated using different
models, which has been presented in Figure~\ref{PC_Comp}. It has been
observed that generally at lower temperatures, the IPM and LSSM computed PC
rates due to $^{45-50,55}$V isotopes are larger than our rates. However at
higher temperatures, our rates surpass the LSSM and IPM rates. From this
list, the comparison results for one even-A ($^{48}$V) and two odd-A
($^{45}$V and $^{49}$V) nuclei are shown in Figure~\ref{PC_Comp}. For example, in case of
$^{48}$V, for temperatures (T$_{9} < 5$), PC rates computed using LSSM and
IPM are greater than our estimated rates by about 1-8 orders of magnitude. Lower the
temperature, larger is the difference between the rates. As the temperature
rises, the difference between the rates reduces and for temperatures
T$_{9} \ge 5$, our rates previal the IPM and LSSM rates by 1-2 orders of
magnitude. For $^{45}$V, LSSM and IPM rates are bigger than pn-QRPA
rates by about a factor two- to three orders of magnitude at temperatures T$_{9} \le
10$. At T$_{9} = 30$, QRPA rates exceed the LSSM and IPM rates by factor 2.
For the next three cases $^{51,52,54}$V, at lower temperatures and densities,
there is a reasonable comparison between different model calculated rates. At
higher temperature QRPA rates get bigger by some factors to an order of
magnitude (see for example the results for $^{51}$V and $^{54}$V). For $^{56-58}$V, QRPA rates in general exceed the
corresponding IPM and LSSM rates at all densities and temperatures by factor of
two to an order of magnitude. The results for one such case ($^{58}$V) are
shown in Figure~\ref{PC_Comp}. These differences in the PC rates calculated
using various models can again be due to the earlier stated reasons. The
core-collapse simulators should take notice of the enhancement in our
computed LC rates at higher temperatures.

\section{Conclusions}
\label{sec:conclusions}

We have calculated the lepton capture rates of twenty two isotopes of vanadium using the
deformed pn-QRPA theory. The calculations of these rates are based on
the GT strength distributions previously reported in Ref.~\citep{shehzadi20}.
The rates were evaluated over a broad temperature ($10^{7} - 3 \times
10^{10}$)\;K and density ($10-10^{11}$)\;g/cm$^{3}$ domain. It has been
observed that, for lighter V isotopes having $43 \le A \le 50$, EC rates are
bigger as compared to the corresponding PC rates. In case of heavier isotopes with $A >
50$, the PC rates are bigger in comparison to the EC rates at lower temperatures and
densities ($\log\rho \text{Y}_{e} < 11$). However, for $^{51-57}$V, the EC
rates become comparable to PC rates at higher temperatures. At high densities, the PC rates are insignificant in comparison to EC rates.

Our calculated lepton capture rates were also compared with the previously
reported LSSM and IPM calculated capture rates. It has been observed that at
higher stellar temperatures, the pn-QRPA computed LC rates are
enhanced as compared to the corresponding LSSM and IPM rates by up to 1-2
orders of magnitude. One of the primary reasons which could cause these
differences is the use of Brink hypothesis by IPM and LSSM calculations,
which is considered a poor approximation in the estimation of GT transition strength from excited states~\citep{Mis14,Joh15}. Our model, on the other hand, treats contribution
from excited states microscopically. A larger model space up to 7 major
oscillator shells was employed in pn-QRPA calculations. GT strength quenching,
and wrong placing of GT centroid in IPM calculations and convergence issues
faced by LSSM calculations would also contribute to the differences observed
in their and our reported capture rates. From simulation studies, the
enhanced LC rates may have an impact on the shock wave energetics
and the late stellar evolution phases. We encourage the core collapse
simulators to test run their stellar codes to analyze the effect of our enhanced
LC rates.

\acknowledgments J.-U. Nabi would like to
acknowledge the support of the Higher Education Commission Pakistan
through project numbers 5557/KPK\\/NRPU/R$\&$D/HEC/2016,
9-5(Ph-1-MG-7)/PAK-\\TURK/R$\&$D/HEC/2017.

\clearpage \onecolumn

\begin{table}[pt]
\caption{\small The pn-QRPA calculated lepton capture rates on
$^{43-46}$V isotopes at various selected densities and temperatures
in stellar environment. $\log\rho \text{Y}_{e}$ has units of
g/cm$^{3}$, where $\rho$ is the baryon density and Y$_{e}$ is the
ratio of the lepton number to the baryon number. Temperature
(T$_{9}$) is given in units of $10^{9}$\;K. $\lambda^{\text{EC}}$
($\lambda^{\text{PC}}$) are the weak decay rates as a result of
electron capture and positron capture, respectively and are given in
units of s$^{-1}$.}\label{V43-46} \centering {\small
\begin{tabular}{cccccccccccc}
& & & & & & & & &\\
\toprule \multirow{2}{*}{$\log\rho$Y$_{e}$} &
\multirow{2}{*}{T$_{9}$} & \multicolumn{2}{c}{$^{43}$V}&
\multicolumn{2}{c}{$^{44}$V} & \multicolumn{2}{c}{$^{45}$V}&
\multicolumn{2}{c}{$^{46}$V}\\
\cmidrule{3-4}  \cmidrule{5-6}  \cmidrule{7-8} \cmidrule{9-10}& &
\multicolumn{1}{c}{$\lambda^{\text{EC}}$} &
\multicolumn{1}{c}{$\lambda^{\text{PC}}$} &
\multicolumn{1}{c}{$\lambda^{\text{EC}}$} &
\multicolumn{1}{c}{$\lambda^{\text{PC}}$} &
\multicolumn{1}{c}{$\lambda^{\text{EC}}$} &
\multicolumn{1}{c}{$\lambda^{\text{PC}}$} &
\multicolumn{1}{c}{$\lambda^{\text{EC}}$} &
\multicolumn{1}{c}{$\lambda^{\text{PC}}$} \\
\midrule
3 & 1 & 1.77E-04 & 4.89E-85 & 1.61E-04 & 5.36E-78 & 8.87E-05 & 3.64E-68 & 1.84E-03 & 1.39E-60\tabularnewline
3 & 3 & 4.46E-02 & 5.68E-29 & 3.77E-02 & 1.48E-26 & 1.99E-02 & 5.07E-24 & 5.68E-01 & 2.43E-19\tabularnewline
3 & 5 & 3.15E-01 & 1.44E-17 & 3.34E-01 & 6.31E-16 & 1.44E-01 & 7.23E-15 & 4.57E+00 & 6.24E-11\tabularnewline
3 & 10 & 4.02E+00 & 1.35E-08 & 6.46E+00 & 2.09E-07 & 2.10E+00 & 3.00E-07 & 6.64E+01 & 2.61E-04\tabularnewline
3 & 30 & 1.05E+03 & 6.08E-01 & 2.80E+03 & 4.67E+00 & 7.71E+02 & 2.24E+00 & 5.31E+03 & 5.13E+01\tabularnewline
 &  &  &  &  &  &  &  &  & \tabularnewline
7 & 1 & 9.02E-01 & 7.40E-91 & 8.34E-01 & 8.11E-84 & 4.72E-01 & 5.51E-74 & 1.06E+01 & 2.10E-66\tabularnewline
7 & 3 & 1.17E+00 & 1.09E-30 & 9.93E-01 & 2.84E-28 & 5.33E-01 & 9.77E-26 & 1.58E+01 & 4.68E-21\tabularnewline
7 & 5 & 1.34E+00 & 2.84E-18 & 1.43E+00 & 1.25E-16 & 6.19E-01 & 1.43E-15 & 2.00E+01 & 1.23E-11\tabularnewline
7 & 10 & 4.97E+00 & 1.07E-08 & 7.98E+00 & 1.67E-07 & 2.61E+00 & 2.39E-07 & 8.22E+01 & 2.08E-04\tabularnewline
7 & 30 & 1.06E+03 & 6.04E-01 & 2.82E+03 & 4.62E+00 & 7.76E+02 & 2.22E+00 & 5.35E+03 & 5.09E+01\tabularnewline
 &  &  &  &  &  &  &  &  & \tabularnewline
11 & 1 & 1.10E+05 & 1.00E-100 & 9.48E+04 & 1.00E-100 & 1.07E+05 & 1.00E-100 & 2.72E+06 & 1.00E-100\tabularnewline
11 & 3 & 1.37E+05 & 3.65E-69 & 1.13E+05 & 9.46E-67 & 1.14E+05 & 3.25E-64 & 3.72E+06 & 1.56E-59\tabularnewline
11 & 5 & 1.42E+05 & 1.14E-41 & 1.43E+05 & 5.01E-40 & 1.15E+05 & 5.74E-39 & 3.83E+06 & 4.95E-35\tabularnewline
11 & 10 & 1.50E+05 & 1.31E-20 & 2.12E+05 & 2.04E-19 & 1.24E+05 & 2.92E-19 & 3.55E+06 & 2.55E-16\tabularnewline
11 & 30 & 4.83E+05 & 8.28E-05 & 1.13E+06 & 6.34E-04 & 4.34E+05 & 3.04E-04 & 2.89E+06 & 6.98E-03\tabularnewline
\bottomrule
\end{tabular}}
\end{table}

\begin{table}[pt]
\caption{\small Same as Table~\ref{V43-46}, but for $^{47-50}$V
isotopes .}\label{V47-50} \centering {\small
\begin{tabular}{cccccccccccc}
& & & & & & & & &\\
\toprule \multirow{2}{*}{$\log\rho$Y$_{e}$} &
\multirow{2}{*}{T$_{9}$} & \multicolumn{2}{c}{$^{47}$V}&
\multicolumn{2}{c}{$^{48}$V} & \multicolumn{2}{c}{$^{49}$V}&
\multicolumn{2}{c}{$^{50}$V}\\
\cmidrule{3-4}  \cmidrule{5-6}  \cmidrule{7-8} \cmidrule{9-10}& &
\multicolumn{1}{c}{$\lambda^{\text{EC}}$} &
\multicolumn{1}{c}{$\lambda^{\text{PC}}$} &
\multicolumn{1}{c}{$\lambda^{\text{EC}}$} &
\multicolumn{1}{c}{$\lambda^{\text{PC}}$} &
\multicolumn{1}{c}{$\lambda^{\text{EC}}$} &
\multicolumn{1}{c}{$\lambda^{\text{PC}}$} &
\multicolumn{1}{c}{$\lambda^{\text{EC}}$} &
\multicolumn{1}{c}{$\lambda^{\text{PC}}$} \\
\midrule
3 & 1 & 6.41E-07 & 1.09E-43 & 3.94E-05 & 7.16E-29 & 3.58E-07 & 1.17E-19 & 6.67E-14 & 4.09E-19\tabularnewline
3 & 3 & 1.50E-04 & 3.86E-16 & 6.25E-02 & 4.06E-09 & 2.15E-04 & 5.55E-08 & 1.77E-06 & 4.62E-08\tabularnewline
3 & 5 & 1.34E-03 & 2.34E-10 & 1.03E+00 & 6.21E-05 & 3.45E-03 & 4.42E-05 & 2.30E-04 & 1.99E-05\tabularnewline
3 & 10 & 4.46E-02 & 1.80E-05 & 3.40E+01 & 2.12E-01 & 1.34E-01 & 2.44E-02 & 5.02E-02 & 8.59E-03\tabularnewline
3 & 30 & 1.16E+02 & 3.74E+00 & 6.98E+03 & 5.96E+02 & 1.85E+02 & 4.92E+01 & 3.09E+02 & 6.15E+01\tabularnewline
 &  &  &  &  &  &  &  &  & \tabularnewline
7 & 1 & 3.94E-03 & 1.65E-49 & 4.63E-01 & 1.08E-34 & 4.65E-03 & 1.77E-25 & 2.58E-08 & 6.19E-25\tabularnewline
7 & 3 & 4.32E-03 & 7.45E-18 & 2.19E+00 & 7.82E-11 & 8.02E-03 & 1.07E-09 & 8.47E-05 & 8.89E-10\tabularnewline
7 & 5 & 5.98E-03 & 4.65E-11 & 4.80E+00 & 1.23E-05 & 1.64E-02 & 8.75E-06 & 1.13E-03 & 3.94E-06\tabularnewline
7 & 10 & 5.55E-02 & 1.44E-05 & 4.25E+01 & 1.69E-01 & 1.67E-01 & 1.95E-02 & 6.28E-02 & 6.85E-03\tabularnewline
7 & 30 & 1.17E+02 & 3.72E+00 & 7.03E+03 & 5.90E+02 & 1.86E+02 & 4.88E+01 & 3.11E+02 & 6.10E+01\tabularnewline
 &  &  &  &  &  &  &  &  & \tabularnewline
11 & 1 & 1.40E+04 & 1.00E-100 & 1.27E+06 & 1.00E-100 & 4.09E+04 & 1.00E-100 & 2.40E+04 & 1.00E-100\tabularnewline
11 & 3 & 1.49E+04 & 2.48E-56 & 3.50E+06 & 2.61E-49 & 4.01E+04 & 3.56E-48 & 4.01E+04 & 2.96E-48\tabularnewline
11 & 5 & 1.53E+04 & 1.87E-34 & 4.23E+06 & 4.93E-29 & 3.92E+04 & 3.52E-29 & 5.35E+04 & 1.58E-29\tabularnewline
11 & 10 & 1.73E+04 & 1.76E-17 & 4.46E+06 & 2.07E-13 & 3.85E+04 & 2.38E-14 & 7.96E+04 & 8.47E-15\tabularnewline
11 & 30 & 9.86E+04 & 5.11E-04 & 4.63E+06 & 8.17E-02 & 1.84E+05 & 6.76E-03 & 3.56E+05 & 8.47E-03\tabularnewline
 \bottomrule
\end{tabular}}
\end{table}

\begin{table}[pt]
\caption{\small Same as Table~\ref{V43-46}, but for $^{51-54}$V
isotopes .}\label{V51-54} \centering {\small
\begin{tabular}{cccccccccccc}
& & & & & & & & &\\
\toprule \multirow{2}{*}{$\log\rho$Y$_{e}$} &
\multirow{2}{*}{T$_{9}$} & \multicolumn{2}{c}{$^{51}$V}&
\multicolumn{2}{c}{$^{52}$V} & \multicolumn{2}{c}{$^{53}$V}&
\multicolumn{2}{c}{$^{54}$V}\\
\cmidrule{3-4}  \cmidrule{5-6}  \cmidrule{7-8} \cmidrule{9-10}& &
\multicolumn{1}{c}{$\lambda^{\text{EC}}$} &
\multicolumn{1}{c}{$\lambda^{\text{PC}}$} &
\multicolumn{1}{c}{$\lambda^{\text{EC}}$} &
\multicolumn{1}{c}{$\lambda^{\text{PC}}$} &
\multicolumn{1}{c}{$\lambda^{\text{EC}}$} &
\multicolumn{1}{c}{$\lambda^{\text{PC}}$} &
\multicolumn{1}{c}{$\lambda^{\text{EC}}$} &
\multicolumn{1}{c}{$\lambda^{\text{PC}}$} \\
\midrule
3 & 1 & 1.26E-21 & 1.48E-10 & 3.86E-32 & 3.40E-06 & 6.24E-33 & 9.86E-08 & 2.92E-39 & 1.60E-07\tabularnewline
3 & 3 & 1.41E-09 & 2.96E-06 & 9.12E-11 & 8.05E-03 & 6.92E-13 & 1.35E-04 & 1.05E-13 & 3.21E-04\tabularnewline
3 & 5 & 3.30E-06 & 7.78E-05 & 8.51E-06 & 2.27E-01 & 5.09E-08 & 1.54E-03 & 4.60E-08 & 4.85E-03\tabularnewline
3 & 10 & 2.54E-02 & 2.72E-02 & 2.47E-01 & 1.17E+01 & 1.71E-03 & 8.13E-02 & 3.01E-03 & 2.59E-01\tabularnewline
3 & 30 & 4.98E+02 & 2.79E+02 & 3.72E+03 & 3.98E+03 & 1.31E+02 & 2.82E+02 & 2.14E+02 & 5.21E+02\tabularnewline
 &  &  &  &  &  &  &  &  & \tabularnewline
7 & 1 & 8.30E-16 & 2.23E-16 & 2.55E-26 & 5.16E-12 & 4.08E-27 & 1.49E-13 & 1.93E-33 & 2.43E-13\tabularnewline
7 & 3 & 6.34E-08 & 5.75E-08 & 4.73E-09 & 1.58E-04 & 3.54E-11 & 2.64E-06 & 5.45E-12 & 6.32E-06\tabularnewline
7 & 5 & 1.55E-05 & 1.56E-05 & 4.31E-05 & 4.59E-02 & 2.56E-07 & 3.13E-04 & 2.32E-07 & 9.86E-04\tabularnewline
7 & 10 & 3.18E-02 & 2.18E-02 & 3.10E-01 & 9.40E+00 & 2.15E-03 & 6.50E-02 & 3.78E-03 & 2.08E-01\tabularnewline
7 & 30 & 5.01E+02 & 2.77E+02 & 3.76E+03 & 3.95E+03 & 1.32E+02 & 2.81E+02 & 2.16E+02 & 5.18E+02\tabularnewline
 &  &  &  &  &  &  &  &  & \tabularnewline
11 & 1 & 2.62E+04 & 1.00E-100 & 4.93E+06 & 1.00E-100 & 2.84E+04 & 1.00E-100 & 2.82E+04 & 1.00E-100\tabularnewline
11 & 3 & 3.64E+04 & 1.92E-46 & 3.62E+06 & 5.26E-43 & 2.91E+04 & 8.81E-45 & 3.80E+04 & 2.11E-44\tabularnewline
11 & 5 & 3.98E+04 & 6.27E-29 & 3.30E+06 & 1.86E-25 & 2.99E+04 & 1.26E-27 & 4.66E+04 & 3.99E-27\tabularnewline
11 & 10 & 5.19E+04 & 2.72E-14 & 3.40E+06 & 1.18E-11 & 3.43E+04 & 8.09E-14 & 6.81E+04 & 2.61E-13\tabularnewline
11 & 30 & 4.73E+05 & 3.88E-02 & 5.27E+06 & 5.60E-01 & 2.00E+05 & 3.92E-02 & 3.48E+05 & 7.24E-02\tabularnewline
 \bottomrule
\end{tabular}}
\end{table}

\begin{table}[pt]
\caption{\small Same as Table~\ref{V43-46}, but for $^{55-58}$V
isotopes .}\label{V55-58} \centering {\small
\begin{tabular}{cccccccccccc}
& & & & & & & & \\
\toprule \multirow{2}{*}{$\log\rho$Y$_{e}$} &
\multirow{2}{*}{T$_{9}$} & \multicolumn{2}{c}{$^{55}$V}&
\multicolumn{2}{c}{$^{56}$V} & \multicolumn{2}{c}{$^{57}$V} & \multicolumn{2}{c}{$^{58}$V}\\
\cmidrule{3-4}  \cmidrule{5-6}  \cmidrule{7-8} \cmidrule{9-10}& &
\multicolumn{1}{c}{$\lambda^{\text{EC}}$} &
\multicolumn{1}{c}{$\lambda^{\text{PC}}$} &
\multicolumn{1}{c}{$\lambda^{\text{EC}}$} &
\multicolumn{1}{c}{$\lambda^{\text{PC}}$} &
\multicolumn{1}{c}{$\lambda^{\text{EC}}$} &
\multicolumn{1}{c}{$\lambda^{\text{PC}}$} &
\multicolumn{1}{c}{$\lambda^{\text{EC}}$} &
\multicolumn{1}{c}{$\lambda^{\text{PC}}$} \\
\midrule
3 & 1 & 7.76E-45 & 7.71E-08 & 1.94E-40 & 2.84E-06 & 1.71E-58 & 1.16E-06 & 1.93E-52 & 3.27E-06\tabularnewline
3 & 3 & 3.66E-16 & 7.71E-05 & 4.29E-14 & 2.90E-03 & 1.25E-19 & 1.39E-03 & 1.18E-17 & 9.59E-03\tabularnewline
3 & 5 & 8.41E-10 & 6.01E-04 & 2.96E-08 & 2.77E-02 & 2.87E-11 & 1.30E-02 & 3.30E-10 & 1.09E-01\tabularnewline
3 & 10 & 3.05E-04 & 1.53E-02 & 3.10E-03 & 8.09E-01 & 2.37E-04 & 5.42E-01 & 6.82E-04 & 2.40E+00\tabularnewline
3 & 30 & 1.20E+02 & 4.66E+01 & 3.09E+02 & 7.29E+02 & 1.76E+02 & 8.79E+02 & 3.56E+02 & 1.55E+03\tabularnewline
 &  &  &  &  &  &  &  &  & \tabularnewline
7 & 1 & 5.13E-39 & 1.17E-13 & 1.28E-34 & 4.30E-12 & 1.13E-52 & 1.76E-12 & 1.28E-46 & 4.95E-12\tabularnewline
7 & 3 & 1.90E-14 & 1.52E-06 & 2.23E-12 & 5.71E-05 & 6.50E-18 & 2.74E-05 & 6.12E-16 & 1.89E-04\tabularnewline
7 & 5 & 4.26E-09 & 1.23E-04 & 1.50E-07 & 5.68E-03 & 1.45E-10 & 2.66E-03 & 1.67E-09 & 2.23E-02\tabularnewline
7 & 10 & 3.83E-04 & 1.23E-02 & 3.89E-03 & 6.50E-01 & 2.98E-04 & 4.37E-01 & 8.57E-04 & 1.93E+00\tabularnewline
7 & 30 & 1.21E+02 & 4.62E+01 & 3.12E+02 & 7.24E+02 & 1.77E+02 & 8.73E+02 & 3.59E+02 & 1.54E+03\tabularnewline
 &  &  &  &  &  &  &  &  & \tabularnewline
11 & 1 & 1.04E+03 & 1.00E-100 & 3.27E+03 & 1.00E-100 & 8.28E+02 & 1.00E-100 & 7.14E+02 & 1.00E-100\tabularnewline
11 & 3 & 1.14E+03 & 5.07E-45 & 1.19E+04 & 1.91E-43 & 8.34E+02 & 9.12E-44 & 1.50E+03 & 6.31E-43\tabularnewline
11 & 5 & 1.68E+03 & 5.00E-28 & 2.38E+04 & 2.31E-26 & 9.16E+02 & 1.08E-26 & 2.23E+03 & 9.04E-26\tabularnewline
11 & 10 & 6.71E+03 & 1.57E-14 & 6.11E+04 & 8.30E-13 & 6.25E+03 & 5.57E-13 & 1.44E+04 & 2.47E-12\tabularnewline
11 & 30 & 1.82E+05 & 6.53E-03 & 4.82E+05 & 1.02E-01 & 2.74E+05 & 1.24E-01 & 5.33E+05 & 2.18E-01\tabularnewline
 \bottomrule
\end{tabular}}
\end{table}

\begin{table}[pt]
\caption{\small Same as Table~\ref{V43-46}, but for $^{59-61}$V
isotopes .}\label{V59-61} \centering {\small
\begin{tabular}{cccccccccc}
& & & & & & & & &\\
\toprule \multirow{2}{*}{$\log\rho$Y$_{e}$} &
\multirow{2}{*}{T$_{9}$} & \multicolumn{2}{c}{$^{59}$V} &
\multicolumn{2}{c}{$^{60}$V}&
\multicolumn{2}{c}{$^{61}$V}\\
\cmidrule{3-4}  \cmidrule{5-6}  \cmidrule{7-8} & &
\multicolumn{1}{c}{$\lambda^{\text{EC}}$} &
\multicolumn{1}{c}{$\lambda^{\text{PC}}$} &
\multicolumn{1}{c}{$\lambda^{\text{EC}}$} &
\multicolumn{1}{c}{$\lambda^{\text{PC}}$} &
\multicolumn{1}{c}{$\lambda^{\text{EC}}$} &
\multicolumn{1}{c}{$\lambda^{\text{PC}}$} \\
\midrule
3 & 1 & 2.94E-67 & 4.07E-06 & 3.01E-59 & 1.46E-06 & 9.38E-76 & 3.48E-06\tabularnewline
3 & 3 & 3.54E-22 & 4.86E-03 & 2.50E-19 & 2.76E-03 & 8.15E-25 & 4.14E-03\tabularnewline
3 & 5 & 1.33E-12 & 4.56E-02 & 8.28E-11 & 2.96E-02 & 4.36E-14 & 3.58E-02\tabularnewline
3 & 10 & 8.34E-05 & 2.59E+00 & 7.40E-04 & 6.08E-01 & 1.77E-05 & 6.37E-01\tabularnewline
3 & 30 & 2.14E+02 & 4.00E+03 & 4.70E+02 & 3.44E+02 & 9.29E+01 & 3.52E+02\tabularnewline
 &  &  &  &  &  &  & \tabularnewline
7 & 1 & 1.94E-61 & 6.17E-12 & 1.99E-53 & 2.21E-12 & 6.19E-70 & 5.28E-12\tabularnewline
7 & 3 & 1.84E-20 & 9.59E-05 & 1.30E-17 & 5.45E-05 & 4.24E-23 & 8.18E-05\tabularnewline
7 & 5 & 6.70E-12 & 9.33E-03 & 4.19E-10 & 6.08E-03 & 2.20E-13 & 7.36E-03\tabularnewline
7 & 10 & 1.05E-04 & 2.08E+00 & 9.29E-04 & 4.90E-01 & 2.23E-05 & 5.14E-01\tabularnewline
7 & 30 & 2.16E+02 & 3.97E+03 & 4.73E+02 & 3.41E+02 & 9.38E+01 & 3.49E+02\tabularnewline
 &  &  &  &  &  &  & \tabularnewline
11 & 1 & 4.16E+02 & 1.00E-100 & 1.04E+02 & 1.00E-100 & 8.04E+01 & 1.00E-100\tabularnewline
11 & 3 & 5.06E+02 & 3.20E-43 & 2.51E+02 & 1.82E-43 & 1.19E+02 & 2.72E-43\tabularnewline
11 & 5 & 6.25E+02 & 3.79E-26 & 4.33E+02 & 2.47E-26 & 1.50E+02 & 2.99E-26\tabularnewline
11 & 10 & 5.55E+03 & 2.68E-12 & 1.18E+04 & 6.30E-13 & 4.81E+02 & 6.64E-13\tabularnewline
11 & 30 & 3.96E+05 & 5.73E-01 & 6.73E+05 & 4.86E-02 & 1.34E+05 & 5.06E-02\tabularnewline
 \bottomrule
\end{tabular}}
\end{table}

\begin{table}[pt]
\caption{\small Same as Table~\ref{V43-46}, but for $^{62-64}$V
isotopes .}\label{V62-64} \centering {\small
\begin{tabular}{cccccccccc}
& & & & & & & & \\
\toprule \multirow{2}{*}{$\log\rho$Y$_{e}$} &
\multirow{2}{*}{T$_{9}$} & \multicolumn{2}{c}{$^{62}$V}&
\multicolumn{2}{c}{$^{63}$V} & \multicolumn{2}{c}{$^{64}$V}\\
\cmidrule{3-4}  \cmidrule{5-6}  \cmidrule{7-8} & &
\multicolumn{1}{c}{$\lambda^{\text{EC}}$} &
\multicolumn{1}{c}{$\lambda^{\text{PC}}$} &
\multicolumn{1}{c}{$\lambda^{\text{EC}}$} &
\multicolumn{1}{c}{$\lambda^{\text{PC}}$} &
\multicolumn{1}{c}{$\lambda^{\text{EC}}$} &
\multicolumn{1}{c}{$\lambda^{\text{PC}}$} \\
\midrule
3 & 1 & 4.35E-69 & 9.38E-06 & 5.06E-85 & 1.06E-05 & 8.47E-81 & 1.28E-05\tabularnewline
3 & 3 & 1.73E-22 & 1.65E-02 & 1.11E-27 & 1.40E-02 & 4.88E-26 & 3.55E-02\tabularnewline
3 & 5 & 1.34E-12 & 1.48E-01 & 9.23E-16 & 1.17E-01 & 1.23E-14 & 3.51E-01\tabularnewline
3 & 10 & 1.34E-04 & 2.65E+00 & 2.27E-06 & 2.14E+00 & 1.16E-05 & 5.71E+00\tabularnewline
3 & 30 & 2.48E+02 & 1.29E+03 & 2.82E+01 & 1.59E+03 & 8.32E+01 & 1.94E+03\tabularnewline
 &  &  &  &  &  &  & \tabularnewline
7 & 1 & 2.87E-63 & 1.42E-11 & 3.34E-79 & 1.60E-11 & 5.60E-75 & 1.94E-11\tabularnewline
7 & 3 & 9.02E-21 & 3.26E-04 & 5.74E-26 & 2.77E-04 & 2.53E-24 & 7.01E-04\tabularnewline
7 & 5 & 6.79E-12 & 3.03E-02 & 4.67E-15 & 2.40E-02 & 6.24E-14 & 7.23E-02\tabularnewline
7 & 10 & 1.67E-04 & 2.14E+00 & 2.85E-06 & 1.73E+00 & 1.46E-05 & 4.61E+00\tabularnewline
7 & 30 & 2.49E+02 & 1.28E+03 & 2.84E+01 & 1.57E+03 & 8.38E+01 & 1.92E+03\tabularnewline
 &  &  &  &  &  &  & \tabularnewline
11 & 1 & 2.66E+02 & 1.00E-100 & 8.63E-01 & 1.00E-100 & 3.52E+00 & 1.00E-100\tabularnewline
11 & 3 & 4.29E+02 & 1.08E-42 & 1.07E+00 & 9.23E-43 & 1.64E+01 & 2.33E-42\tabularnewline
11 & 5 & 4.80E+02 & 1.23E-25 & 1.28E+00 & 9.77E-26 & 2.43E+01 & 2.93E-25\tabularnewline
11 & 10 & 2.58E+03 & 2.75E-12 & 4.83E+01 & 2.23E-12 & 2.13E+02 & 5.96E-12\tabularnewline
11 & 30 & 3.53E+05 & 1.82E-01 & 4.03E+04 & 2.24E-01 & 1.18E+05 & 2.77E-01\tabularnewline
 \bottomrule
\end{tabular}}
\end{table}

\end{document}